\documentclass[11pt,showpacs,preprintnumbers,amsmath,amssymb,prd,nofootinbib,superscriptaddress]{revtex4-2}

\usepackage{dcolumn}
\usepackage{bm}
\usepackage{ifpdf}
\usepackage{hyperref}
\usepackage{bm}
\usepackage{xcolor,color,graphicx,graphics,physics}
\usepackage[spanish,english]{babel}
\usepackage[latin1]{inputenc}
\usepackage[OT1]{fontenc}
\usepackage{latexsym,amssymb,amsmath,amsfonts, slashed}
\usepackage{makeidx}
\usepackage{epsfig,subfigure}
\usepackage{natbib}
\usepackage{epstopdf}
\usepackage{mathrsfs}
\usepackage{hyperref}
\hypersetup{colorlinks=true, linkcolor=blue, citecolor=blue, urlcolor=blue}
\usepackage{enumerate}

\usepackage{fixmath}

\everymath{\displaystyle}
\usepackage{graphicx}

\usepackage[T1]{fontenc}
\usepackage{amsmath}
\usepackage{amssymb}
\usepackage{graphicx}
\usepackage{xcolor}

\newcommand{\bea}{\begin{eqnarray}}
\newcommand{\eea}{\end{eqnarray}}

\newcommand{\orcid}[1]{\href{https://orcid.org/#1}{\includegraphics[width=10pt]{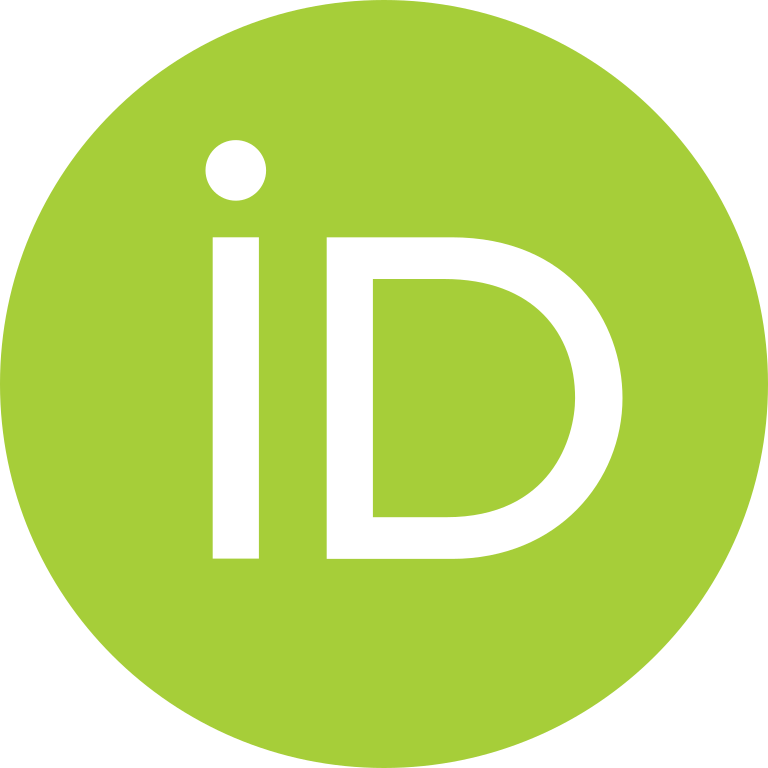}}}

\begin{document}

\title{G\"{o}del-type universes in energy-momentum-squared gravity}

\author{\'A. J. C. Canuto  \orcid{0009-0001-9886-9366}}
\email{alefcanuto.fis@fisica.ufmt.br}
\affiliation{Instituto de F\'{\i}sica, Universidade Federal de Mato Grosso,\\
78060-900, Cuiab\'{a}, Mato Grosso, Brazil}

\author{A. F. Santos \orcid{0000-0002-2505-5273}}
\email{alesandroferreira@fisica.ufmt.br}
\affiliation{Instituto de F\'{\i}sica, Universidade Federal de Mato Grosso,\\
78060-900, Cuiab\'{a}, Mato Grosso, Brazil}

\begin{abstract}

In this paper, a modification of general relativity is considered. It consists of generalizing the Lagrangian of matter in a non-linear way, that is, replacing the curvature scalar $R$ by a function $f(R,T_{\mu\nu} T^{\mu\nu} )$,  where $T_{\mu\nu}$ is the energy-momentum tensor. The main objective is to investigate the issue of causality in this gravitational model. To study the causality and/or its violation the G\"{o}del-type solutions are used. For such development, different matter contents are chosen. A critical radius, beyond which causality is violated, is calculated. It is shown that both causal and non-causal solutions are allowed.

\end{abstract}

\maketitle

\section{Introduction}

General Relativity (GR) or Einstein's theory of gravitation \cite{1916A,einstein1915field}, is the best theory of gravitation for describing large-scale interactions. The GR has been successfully tested through several observational results, such as the predicted value for the advance of Mercury's perihelion, the bending of light around the Sun, gravitational redshift, detection of gravitational waves, among others \cite{review1, review2}. It is clear that GR is an established theory, however some challenges need to be resolved. The most important challenges are: (i) it is a classical theory, i.e., there is no quantum version of GR as there is for the other fundamental forces of nature and (ii) the accelerated expansion of the universe, confirmed by various observational tests \cite{Riess_1998, Perlmutter_1999, Riess_2004}. Since GR does not explain these problems, alternative theories are constructed and investigated.

Over the years, some models have been proposed to try to explain the accelerated expansion of the universe. There are two different ways to develop these theories, one adds an exotic component of energy, called dark energy \cite{Carroll_2001, quinte, Capozziello_2006, taquions, gasChaplygin}, and the other modifies or generalizes the Einstein-Hilbert action. Such modifications change the geometry or matter or both parts of the Einstein field equations. The simplest extension of GR is obtained by replacing the Ricci scalar $R$ in the Einstein-Hilbert action with a general function of $R$, leading to $f(R)$ theories of gravity \cite{f(R), Nojiri, Carroll}. Another interesting generalization is the $f(R, T)$ gravity, where the gravitational action is an arbitrary function of the
Ricci scalar $R$ and of the trace of the energy-momentum tensor $T$ \citep{Harko}. It is important to note that $f(R)$ and $f(R,T)$ are two of several models of modified gravity theory. For a review of modified gravity theories, see \cite{Clifton, Shan}.

In this work, the $f(R,T^2)$ gravity is considered. This is a theory where the gravitational Lagrangian depends on the Ricci scalar $R$ and on the contraction of the energy-momentum tensor with itself, i.e., $T^2=T_{\mu\nu}T^{\mu\nu}$ \cite{Kat, Roshan, Barrow}. It should be noted that the field equations differ from GR only in the presence of matter sources. The term  $f(T_{\mu\nu}T^{\mu\nu})$ can be introduced in different ways, which leads to different versions of the theory. For example, the version with $f(T_{\mu\nu}T^{\mu\nu})\propto(T_{\mu\nu}T^{\mu\nu})^\eta$ is known as energy-momentum powered gravity, where $\eta$ is a constant. While the model with $f(T_{\mu\nu}T^{\mu\nu})\propto\ln(T_{\mu\nu}T^{\mu\nu})$ is called energy-momentum log gravity. This theory has received a lot of attention and has been studied in different contexts. In \cite{Cosmic} the cosmic acceleration via energy-momentum powered gravity has been studied. Cosmological models including bulk viscous cosmology, loop quantum gravity, k-essence, and brane-world cosmologies in energy-momentum-squared gravity have been discussed \cite{Barrow}. An extension of the standard $\Lambda$CDM model in energy-momentum log gravity has been investigated \cite{Uzun}. Constraints on the energy-momentum squared gravity from neutron stars and its cosmological implications have been considered \cite{Eksi}. Spherically symmetric compact stars have been analyzed \cite{Nari}. Dynamical system analysis for various types of gravity functions $f(R,T^2)$  has been used and the structure of the phase space and the physical implications have been studied \cite{Marciu}. The viability of bouncing cosmology has been explored \cite{Barbar}. A generalized version of energy-momentum squared gravity in the Palatini formalism has been constructed \cite{Palatini}. Quasinormal modes of perturbed black holes have been discussed \cite{Modes}. A particular model of energy-momentum squared gravity, called scale-independent  energy-momentum squared gravity, has been developed and some implications have been considered \cite{scale}. Although the energy-momentum squared gravity has been tested in several contexts, the question of causality and its possible violation has not been developed. Therefore, the main objective of this paper is to address a study on causality.  For such a construction, the G\"{o}del-type solution is considered.

An exact solution of the GR field equations was proposed by Kurt G\"{o}del in 1949 \cite{K.Godel}.  This solution describes a pressureless perfect fluid in rotation, without expansion and exhibits cylindrical symmetry. This cosmological model has as its main characteristic the possibility of Closed Time-like Curves (CTCs). These CTCs allow an observer to go back into the past, which leads to the violation of causality. However, it is important to note that in GR the space-times locally have the same causal structure of the special relativity. But on a non-local scale, the causality can be violated. CTCs is not a unique feature of the G\"{o}del metric, there are other cosmological models that exhibit similar curves, such as van Stockum cylinder \cite{Stockum}, Kerr space-time \cite{Kerr}, cosmic string \cite{string}, among others. In order to calculate a critical radius, beyond which the causality is violated, a generalization of the G\"{o}del metric, called G\"{o}del-type metric \cite{Godel-type.Reboucas}, is considered. In this context, let's verify the consistency of the field equations of $f(R,T^2)$ gravity in a cosmological background defined by the G\"{o}del and G\"{o}del-type metrics. Considering different content of matter, causal and non-causal regions arise in this gravitational model.

The present paper is organized as follows. In section II, an introduction to the $f(R,T^2)$ gravity is made. In section III, the G\"{o}del metric is introduced and the field equations are solved. The causality problem is discussed. In section IV, the G\"{o}del-type metric is considered. First, a perfect fluid is chosen as the content of matter. In this case, the causality is naturally violated. A critical radius, which defines causal and non-causal regions, is calculated. Then, in order to find a causal region, different matter contents are introduced, such as perfect fluid and a scalar field together and only a scalar field. In section V, remarks and conclusions are presented.

\section{$f(R,T^{2})$ Modified Gravity }\label{sec:modified}

In this section, the main objective is to obtain the field equations of  $f(R,T^{2})$ gravity, with $T^2=T_{\mu\nu}T^{\mu\nu}$. The action that describes this theory is,
\begin{equation}\label{eq:1}
    \mathcal{S}=\frac{1}{2\kappa}\int\sqrt{-g}\left(f\left(R,T^{\mu\nu}T_{\mu\nu}\right)+2\Lambda\right)d^{4}x+\int{ \mathcal{L}_{m}\sqrt{-g}d^{4}x},
\end{equation}
where $\kappa=8\pi G$ with $G$ being the gravitational constant, $g$ is the determinant of the metric, $R$ is the Ricci scalar, $\Lambda$ is the cosmological constant,  $\mathcal{L}_m$ is matter Lagrangian and  $T_{\mu\nu}$  is the energy-momentum tensor which is defined as
\begin{equation}\label{eq:2}
     T_{\mu\nu}=-\frac{2}{\sqrt{-g}}\frac{\delta(\sqrt{-g}\mathcal{L}_{m})}{\delta g^{\mu\nu}}.
\end{equation}
Considering that $\mathcal{L}_{m}$ depends only on the metric components, and not on their derivatives, the energy-momentum tensor becomes
\begin{equation}\label{eq:3}
    T_{\mu\nu}=g_{\mu\nu}\mathcal{L}_{m}-2\frac{\partial \mathcal{L}_{m}}{\partial g^{\mu\nu}}.
\end{equation}

The variation of Eq. \eqref{eq:1} leads to
\begin{equation}\label{eq:4}
    \delta\mathcal{S}=\frac{1}{2\kappa}\int \sqrt{-g} \left\{ f_{R}\delta R+f_{T^{2}}\delta T^2-\frac{1}{2}g_{\mu\nu}f\delta g^{\mu\nu}-g_{\mu\nu}\Lambda\delta g^{\mu\nu}+\frac{1}{\sqrt{-g}}\delta\left(\sqrt{-g}\mathcal{L}_{m}\right)\right\} d^{4}x.
\end{equation}
Here, for simplicity,  have been defined $f=f(R,T_{\mu\nu}T^{\mu\nu})$,  $f_R=\frac{\partial f}{\partial R}$ and $f_{T^2}=\frac{\partial f}{\partial T^2}$. The variation of the Ricci scalar is known and gives the result
\bea
\delta R=R_{\mu\nu}\delta g^{\mu\nu}+g_{\mu\nu}\Box\delta g^{\mu\nu}-\nabla_\mu\nabla_\nu\delta g^{\mu\nu}.
\eea
Taking the variation of $T^2$ with respect to the metric we obtain
\bea
 \Theta_{\mu\nu}\equiv\frac{\delta (T_{\alpha\beta}T^{\alpha\beta})}{\delta g^{\mu\nu}}&=& \frac{(\delta T_{\alpha\beta})}{\delta g^{\mu\nu}}T^{\alpha\beta} + T_{\alpha\beta}\frac{(\delta T^{\alpha\beta})}{\delta g^{\mu\nu}}\nonumber\\
&=& \frac{(\delta T_{\alpha\beta})}{\delta g^{\mu\nu}}T^{\alpha\beta} + 2T_{\mu}^{\alpha}T_{\nu\alpha}+T^{\alpha\beta}\frac{\delta T_{\alpha\beta}}{\delta g^{\mu\nu}},\label{6}
\eea
Using Eq. \eqref{eq:3}, the variation in the energy-momentum tensor given in Eq. (\ref{6}) is written as
\begin{equation}\label{eq:10}
    \frac{(\delta T_{\alpha\beta})}{\delta g^{\mu\nu}}T^{\alpha\beta}=-\mathcal{L}_{m}T_{\mu\nu}-\frac{1}{2}TT_{\mu\nu}+\frac{1}{2}g_{\mu\nu}\mathcal{L}_{m}T-2\frac{\partial^{2} \mathcal{L}_{m}}{\partial g^{\mu\nu} \partial g^{\alpha\beta}}T^{\alpha\beta}.
\end{equation}
With this result, Eq. (\ref{6}) becomes
\begin{equation}\label{eq:11}
    \Theta_{\mu\nu}= -2\mathcal{L}_{m}\left(T_{\mu\nu}-\frac{1}{2}g_{\mu\nu}T \right)-TT_{\mu\nu} + 2T_{\mu}^{\alpha}T_{\nu\alpha}-4T^{\alpha\beta}\frac{\partial^{2} \mathcal{L}_{m}}{\partial g^{\mu\nu} \partial g^{\alpha\beta}}.
\end{equation}
Thus, returning to Eq. (\ref{eq:4}), the field equations of the $f(R,T^2)$ gravity model are given as
\begin{equation}\label{eq:14}
    f_{R}R_{\mu\nu}-\frac{1}{2}fg_{\mu\nu}-\Lambda g_{\mu\nu}+\left(g_{\mu\nu}\nabla_{\alpha}\nabla^{\alpha}-\nabla_{\mu}\nabla_{\nu}\right)f_{R} =\kappa(T_{\mu\nu}-\frac{1}{\kappa}f_{T^{2}}\Theta_{\mu\nu}).
\end{equation}
Note that for a proper choice of function $f(R, T^2)$ other modified gravities can be obtained, for example, $f=f(R)$ leads to $f(R)$ gravity while $f=f(R,T)$ implies in the field equations of $f(R,T)$ gravity theory.

In the next sections, the field equations (\ref{eq:14}) will be considered to investigate whether this gravitational model allows for the existence of CTCs, which leads to non-causal regions. First, the original G\"{o}del solution is analyzed, and then a generalized model is studied.

\section{G\"{o}del metric in $f(R,T^2)$ gravity}

The G\"{o}del metric was proposed by mathematician and logician Kurt G\"{o}del in 1949 \cite{K.Godel}. It is an exact solution of the field equations of general relativity that allows the existence of Closed Time-like Curves (CTCs). This means that an observer moving on a CTC can find a point in the future that is also his starting point in the past. It is worth noting that the existence of CTCs in the G\"{o}del metric is considered paradoxical,  as it allows past and future events to interact. However, the exact solution of the G\"{o}del metric is interesting from a theoretical point of view and has been studied in several areas of physics and mathematics. The line element that describes this universe is given as
\begin{equation}\label{eq:15}
		ds^{2}=a^{2}\left(dt^{2}-dx^{2}+\frac{e^{2x}}{2}dy^{2}-dz^{2}+2e^{x}dtdy\right),
\end{equation}
where $a$ is a positive constant. The main ingredients associated with Eq. (\ref{eq:15}) for working with Eq. (\ref{eq:14}) are:  (i) the components of metric and its inverse
\begin{equation}\label{eq:16}
    g_{\mu\nu}=a^{2}\left(\begin{array}{cccc}
1 & 0 & e^{x} & 0\\
0 & -1 & 0 & 0\\
e^{x} & 0 & \frac{e^{2x}}{2} & 0\\
0 & 0 & 0 & -1
\end{array}\right),
\end{equation}
\begin{equation}\label{eq:17}
    g^{\mu\nu}=\frac{1}{a^{2}}\left(\begin{array}{cccc}
-1 & 0 & 2e^{-x} & 0\\
0 & -1 & 0 & 0\\
2e^{-x} & 0 & -2e^{-2x} & 0\\
0 & 0 & 0 & -1
\end{array}\right);
\end{equation}
(ii) the non-zero Ricci tensor components
\begin{equation}\label{eq:18}
    R_{00}=1;\,\,R_{02}=e^{x};\,\,R_{22}=e^{2x},
\end{equation} 
and (iii) Ricci scalar
\begin{equation}\label{eq:19}
    R=\frac{1}{a^2}.
\end{equation}

In addition to these geometry elements, a matter content must be considered. Here, a perfect fluid is chosen, whose energy-momentum tensor is defined as
\begin{equation}\label{eq:26}
    T_{\mu\nu}=(\rho+p)u_{\mu}u_{\nu} - pg_{\mu\nu},
\end{equation}
where $\rho$ is the energy density, $p$ is the pressure and $u$ is a unit time-like vector whose covariant components are  $u_\mu=(a,0,ae^x,0)$. Explicitly, the components of the energy-momentum tensor are given as
\begin{equation}\label{eq:27}
    T_{\mu\nu}=\left(\begin{array}{cccc}
\rho a^{2} & 0 & \rho a^{2}e^{x} & 0\\
0 & pa^{2} & 0 & 0\\
\rho a^{2}e^{x} & 0 & \left(\rho+\frac{p}{2}\right)a^{2}e^{2x} & 0\\
0 & 0 & 0 & 0
\end{array}\right).
\end{equation}

Taking the Lagrangian $\mathcal{L}_{m}=-p$, which describes a perfect fluid, and the energy-momentum tensor defined in Eq. (\ref{eq:26}), the tensor $ \Theta_{\mu\nu}$ given in Eq.(\ref{eq:11}) can be written as
\begin{equation}\label{eq:28}
    \varTheta_{\mu\nu}=-(\rho^{2}-p^{2})u_{\mu}u_{\nu},
\end{equation}
whose components are
\begin{equation}\label{eq:29}
 \varTheta_{\mu\nu}=\left(\begin{array}{cccc}
-(\rho^{2}-p^{2})a^{2} & 0 & -(\rho^{2}-p^{2})a^{2}e^{x} & 0\\
0 & 0 & 0 & 0\\
-(\rho^{2}-p^{2})a^{2}e^{x} & 0 & -(\rho^{2}-p^{2})a^{2}e^{2x} & 0\\
0 & 0 & 0 & 0
\end{array}\right).
\end{equation}

Now, let's write the set of field equations. However, it is important to note that the Ricci scalar is a constant value, which implies that the field equation (\ref{eq:14}) reduces to
\begin{equation}\label{eq:31}
    f_{R}R_{\mu\nu}-\frac{1}{2}fg_{\mu\nu}-\Lambda g_{\mu\nu} =\kappa(T_{\mu\nu}-\frac{1}{\kappa}f_{T^{2}}\Theta_{\mu\nu}).
\end{equation}
Consequently, the components of the field equations become
\bea
 \frac{f_{R}}{a^2}-\frac{1}{2}f-\Lambda &=&\kappa\rho + f_{T^{2}}(\rho^2-p^2),\\
 \frac{f_{R}}{a^2}-\frac{1}{4}f-\frac{\Lambda}{2} &=&\kappa(\rho+\frac{p}{2}) + f_{T^{2}}(\rho^2-p^2),\\
  \frac{1}{2}f+\Lambda &=&\kappa p.
\eea
Such a set of equations has the solution
\bea
 \Lambda&=&\frac{1}{2}(2\kappa p-f),\label{cond1} \\
 \rho&=&-\frac{a^2\kappa\pm\sqrt{4a^2f_{R}f_{T^2}+4a^4f_{T^2}^2p^2-4a^4f_{T^2}p\kappa+a^4\kappa^2}}{2a^2f_{T^2}}.  \label{cond2}
\eea
It is important to observe that the $f(R,T^{2})$ modified gravity allows the G\"{o}del solution, then the existence of CTCs is possible and the causality can be violated in this gravitational theory. Although this is the same consequence obtained in general relativity, the conditions (\ref{cond1}) and (\ref{cond2}) for solving the set of equations are different. These conditions are strongly dependent on the function $f(R,T^{2})$ and the matter content. In addition, assuming $f(R,T^{2})=R$ in Eq. (\ref{eq:31}) the results of general relativity are recovered.

In the next section, the G\"{o}del metric is generalized in order to obtain more information about causality violation. To discuss the possibility of finding causal and non-causal regions, different matter contents are considered.

\section{G\"{o}del-type metric in $f(R,T^2)$ modified gravity}

A generalization of the G\"{o}del solution, known as a G\"{o}del-type solution, was developed in \cite{Godel-type.Reboucas}. In this section, the main objective is to investigate the consistency of the G\"{o}del-type metric in $f(R,T^2)$ theory, as well as to analyze the conditions that lead to non-causal regions. Then, if causality is violated, the critical radius will be calculated. In cylindrical coordinates, the G\"{o}del-type metric is given as
\begin{equation}\label{eq:47}
    ds^2=[dt+H(r)d\phi]^{2}-dr^2-D^2(r)d\phi^2-dz^2,
\end{equation}
where the functions $H(r)$ and $D(r)$  must obey the relations
\bea
  \frac{H'(r)}{D(r)}&=&2\omega,\\
   \frac{D''(r)}{D(r)}&=&m^2.
\eea
Here, the prime means derivative with respect to $r$, and $\omega$ and $m$ are free parameters. 

The G\"{o}del-type metrics can be classified into three different classes, i.e., (i) hyperbolic class ($m^2>0$), (ii) trigonometric class ($m^2<0$) and (iii) linear class ($m^2=0$). In this work, only the hyperbolic class is considered.  For this class, the functions $H(r)$ and $D(r)$ are defined as
\bea
H(r)&=&\frac{4\omega}{m^2}sinh^2\left(\frac{mr}{2}\right), \\
D(r) &=&\frac{1}{m}sinh(mr).
\eea

In order to analyze the possibility of occurrence of CTCs, the line element \eqref{eq:47} is written as
\begin{equation}\label{eq:52}
    ds^2=-dt^2-2H(r)dtd\phi+dr^2+G(r)d\phi^2 +dz^2,
\end{equation}
where  $G(r)=D^2(r)-H^2(r)$. The circles defined by $t, z, r = const$, lead to the existence of CTCs when $G(r) < 0$ for a certain range of $r$. In this context, a critical radius, beyond which the causality is violated, is obtained, i.e.,
\begin{equation}\label{eq:53}
    r_{c}=\frac{2}{m}sinh^{-1}\left(\frac{4\omega^2}{m^2}-1\right)^{-1}.
\end{equation}
From the critical radius, it is important to notice that, for $m^2=2\omega^2$ the original G\"{o}del solution is recovered and for  $m^2=4\omega^2$ the critical radius becomes infinite, implying a causal universe. Therefore, for $m^2\geq4\omega^2$ there are no G\"{o}del-type CTCs, and causality violation is avoided.

To solve the field equations more simply, a new basis is chosen \cite{Godel-type.Reboucas}. Then the metric \eqref{eq:52} becomes
\begin{equation}\label{eq:55}
    ds^2=\eta_{AB}\theta^{A}\theta^{B}=\left(\theta^0 \right)^{2}-\left(\theta^1 \right)^{2}-\left(\theta^2 \right)^{2}-\left(\theta^3 \right)^{2},
\end{equation}
with $\eta_{AB}=(+,-,-,-)$ and $  \theta^A=e_{\mu}^{A}dx^{\mu}$. The Latin letters denote the transformed space and $e_{\mu}^{A}$ are the tetrads. The components of $\theta^A$ are
\bea
 \theta^{(0)}&=&dt+H(r)d\phi,\\
 \theta^{(1)}&=& dr,\\
 \theta^{(2)}&=&D(r)d\phi,\\
\theta^{(3)}&=& dz
\eea
and components of the tetrads are
\begin{equation}\label{eq:61}
    e_{0}^{(0)}=1, \, e_{2}^{(0)}=H(r), \,e_{1}^{(1)}=1, \, e_{2}^{(2)}=D(r),\, e_{3}^{(3)}=1.
\end{equation}
The inverse of tetrads, which satisfies the condition $e_{\mu}^{A}e_{B}^{\mu}=\delta_{B}^{A}$, has the following non-zero components
\begin{equation}\label{eq:62}
    e_{(0)}^{0}=e_{(1)}^{1}=e_{(3)}^{3}=1,~~ e_{(2)}^{0}=-\frac{H(r)}{D(r)}, ~~ e_{(2)}^{2}=D^{-1}(r).
\end{equation}

On this new basis, a flat space-time, the non-zero components of the  Ricci tensor are
\bea
 R_{(0)(0)}&=&2\omega^2,\\
 R_{(1)(1)}&=&R_{(2)(2)}=2\omega^2-m^2,
\eea
the scalar curvature is $R=2(m^2-\omega^2)$ and the non-zero components of the Einstein tensor are
\bea
G_{(0)(0)}&=&3\omega^2-m^2,\\
G_{(1)(1)}&=&G_{(2)(2)}=\omega^2,\\
G_{(3)(3)}&=&m^2-\omega^2.
\eea

Now the field equations of $f(R, T^2)$ gravity will be solved for three different matter contents: (i) perfect fluid,  (ii) perfect fluid plus scalar field and (iii) scalar field.

\subsection{Perfect fluid}

In this subsection, the perfect fluid is taken as the content of matter. The energy-momentum tensor describing this matter in flat space-time is given as
\begin{equation}\label{eq:77}
    T_{AB}=\left(\rho+p\right)u_{A}u_{B}-p\eta_{AB},
\end{equation}
where $u_{A}=(1,0,0,0)$ and $ T_{AB}=e_{A}^{\mu}e_{B}^{\nu}T_{\mu\nu}$.  The non-zero components are
\bea
 T_{(0)(0)}=\rho,\quad\quad\quad T_{(1)(1)}= T_{(2)(2)}=  T_{(3)(3)}=p
\eea
and its trace, $T=T_{AB}\eta^{AB}$, is
\begin{equation}\label{eq:83}
    T=\rho-3p.
\end{equation}
In a similar way, the tensor (\ref{eq:28}) in this space-time becomes
\begin{equation}
    \Theta_{AB}=-\left(\rho^2-p^2\right)u_{A}u_{B}.
\end{equation}
 The non-zero component is
\begin{equation}\label{eq:84}
    \Theta_{(0)(0)}=-\left(\rho^2-p^2\right)
\end{equation}
and its trace is given as 
\begin{equation}\label{eq:85}
    \Theta=-\left(\rho^2-p^2\right).
\end{equation}

In local Lorentz co-frame the field equations \eqref{eq:14} becomes
\begin{equation}\label{eq:86}
    f_{R}R_{AB}-\frac{1}{2}f\eta_{AB}-\Lambda \eta_{AB}+\left(\eta_{AB}\nabla_{\alpha}\nabla^{\alpha}-\nabla_{A}\nabla_{B}\right)f_{R} =\kappa(T_{AB}-\frac{1}{\kappa}f_{T^{2}}\Theta_{AB}).
\end{equation}
Considering that the Ricci scalar is a constant and taking the trace of this equation (\ref{eq:86}) as an important constraint, the field equations take the form
\begin{equation}\label{eq:88}
    f_{R}G_{AB}=\kappa T_{AB}-f_{T^2}\Theta_{AB}- \frac{1}{2}\left\{ f+\kappa T-f_{T^{2}}\Theta \right\}\eta_{AB}.
\end{equation}
Then, the field equations for the G\"{o}del-type metric with matter content (\ref{eq:77}) are given by
\bea
2f_{R}\left(3\omega^{2}-m^{2}\right)+F&=&\kappa\left(\rho+3p\right)+f_{T^{2}}\left(\rho^{2}-p^{2}\right),\label{52}\\
2f_{R}\left(\omega^{2}\right)-f&=&\kappa\left(\rho-p\right)+f_{T^{2}}\left(\rho^{2}-p^{2}\right),\label{53}\\  
2f_{R}\left(m^2-\omega^{2}\right)-f&=&\kappa\left(\rho-p\right)+f_{T^{2}}\left(\rho^{2}-p^{2}\right).\label{54}
\eea
Eqs. (\ref{53}) and (\ref{54}) lead to
\begin{equation}\label{eq:92}
    f_{R}\left(2\omega^2-m^2\right)=0.
\end{equation}
Assuming that $f_{R}>0$, Eq. (\ref{eq:92}) gives 
\begin{equation}\label{eq:93} 
    m^2=2\omega^2.
\end{equation}
This condition defines the original G\"{o}del universe. Therefore, the $f{(R,T^2)}$ gravity theory with a perfect fluid as matter content allows the existence of CTCs which implies a violation of the causality. In order to obtain more details about the causality violation, the critical radius is calculated. The remaining field equations imply
\bea
m^2f_{R}+f&=&\kappa\left(\rho+3p\right)+f_{T^{2}}\left(\rho^{2}-p^{2}\right),\\
 m^2f_{R}-f&=&\kappa\left(\rho-p\right)+f_{T^{2}}\left(\rho^{2}-p^{2}\right),
\eea
that lead to
\bea
m=\sqrt{\frac{2\kappa\rho+f+2f_{T^{2}}\left(\rho^{2}-p^{2}\right)}{2f_{R}}}.
\eea
With this result, the critical radius reads
 \begin{equation}\label{eq:98}
     r_{c}=2sinh^{-1}(1)\sqrt{\frac{2f_{R}}{2\kappa\rho+f+2f_{T^{2}}\left(\rho^{2}-p^{2}\right)}}.
 \end{equation}
Therefore, there is a causality violation beyond this critical radius, which explicitly depends on the choice of the function $f(R,T^2)$, its derivatives with respect to $R$ and $T^2$, and the matter content. 

\subsection{Perfect fluid with scalar field}

In the last subsection, it was seen that when the content of matter is just a perfect fluid the causality violation arises naturally. In an attempt to find a causal solution for the G\"{o}del-type metric in $f(R, T^2)$ gravity, let us consider that the matter distribution has two ingredients, a fluid perfect and a scalar field. Then the total energy-momentum tensor is given 
\begin{equation}\label{eq:99}
    T_{AB}=T_{AB}^{(M)} + T_{AB}^{(S)},
\end{equation}
where $T_{AB}^{(M)}$ is the energy-momentum tensor of the perfect fluid, Eq. (\ref{eq:77}), and $T_{AB}^{(S)}$ is the energy-momentum tensor associated with a scalar field, given by
\begin{equation}\label{eq:100}
    T_{AB}^{(S)}= \partial_{A}\phi \partial_{B}\phi - \frac{1}{2}\eta_{AB}\,\eta^{CD} \partial_{C}\phi \partial_{D}\phi.
\end{equation}
For simplicity, and following \cite{Godel-type.Reboucas}, let's assume that the scalar field takes the form $\phi(z) = ez+const$, where $e$ is a constant. Thus, the non-zero components are given as
\begin{equation}\label{eq:112}
    T_{(0)(0)}^{(S)}=-T_{(1)(1)}^{(S)}=-T_{(2)(2)}^{(S)}=T_{(3)(3)}^{(S)}=\frac{e^2}{2}.
\end{equation}

The trace of the total energy-momentum tensor is
\begin{equation}\label{eq:113}
    T=T_{AB}^{(M)}\eta^{AB}+T_{AB}^{(S)}\eta^{AB}=\rho-3p+e^2.
\end{equation}

Considering that the Lagrangian describing the  scalar field is given by
\begin{equation}\label{eq:115}
\mathcal{L}_{m}^{S}=\eta^{MN}\nabla_{M}\phi\nabla_{N}\phi,
\end{equation}
the total tensor $\Theta_{AB}$ can be written as
\begin{equation}\label{eq:114}
\Theta_{AB}=\Theta_{AB}^{(M)}+\Theta_{AB}^{(S)},
\end{equation}
where $\Theta_{AB}^{(M)}$ and $\Theta_{AB}^{(S)}$ are the parts associated with the perfect fluid and the scalar field, respectively. 

Taking these ingredients, the field equations are given as
\bea
2f_{R}\left(3\omega^{2}-m^{2}\right)+f&=&\kappa\left(\rho+3p\right)+f_{T^{2}}\left(\rho^{2}-p^{2}\right),\\
2f_{R}\left(\omega^{2}\right)-f&=&\kappa\left(\rho-p\right)+f_{T^{2}}\left(\rho^{2}-p^{2}\right),\\
2f_{R}\left(m^2-\omega^{2}\right)-f&=&\kappa\left(\rho-p\right)+ 2\kappa e^2+f_{T^{2}}\left(\rho^{2}-p^{2}\right).
\eea
This set of equations can be reduced to
\bea
\kappa e^2&=&f_{R}(m^2-2\omega^2),\\
\kappa p&=&\frac{1}{2}f_{R}(2\omega^2-m^2)+\frac{f}{2},\\
\kappa\rho+f_{T^{2}}\left(\rho^{2}-p^{2}\right)&=&\frac{1}{2}f_{R}(6\omega^2-m^2)-\frac{f}{2}.
\eea
Analyzing these equations, a causal G\"{o}del-type solution arises and is given by
\bea
m^2 &=& 4\omega^2,\\
f_{R}&=&\frac{\kappa e^2}{2\omega^2},\\
\kappa p&=&-\kappa\rho + f_{T ^{2}}\left(\rho^{2}-p^{2}\right)=-\omega^2f_{R}+\frac{f}{2}.
\eea
It is important to emphasize that the condition $m^2= 4\omega^2$ leads to $r_{c}\rightarrow\infty$. Therefore, for this combination of perfect fluid and scalar field as matter content, a causal G\"{o}del-type universe is allowed, i.e., the causality violation is avoided.

\subsection{Scalar field}

In order to find other causal G\"{o}del-type solutions, let's assume that the only source of energy and matter is a scalar field $\phi(z)$. Following the same steps as in the last subsection, it is shown that a unique class of G\"{o}del-type solutions arises, i.e.
\bea
m^2 &=& 4\omega^2,\\
f_{R}&=&\frac{\kappa e^2}{2\omega^2},\\
f&=&\kappa e^2.
\eea
Therefore, this solution leads to an infinite critical radius, as a consequence, causality breaking is not allowed and there are no CTCs in this G\"{o}del-type universe.  A similar analysis has been developed for the $f(R)$ theory \cite{rebouccas2009godel}.

In addition to the investigation developed here, it is important to discuss whether the field equations of $f(R, T^2)$ can have the form of a perfect fluid. This characteristic implies that any source of field equations that can be reformulated into a perfect fluid form is suitable for solving system dynamics. Considering the definition of perfect scalars,  it has been shown that the field equations of some extended theories of gravity contain perfect fluid terms \cite{Capo1, Capo2, Capo3, Sid}. In order to develop such an analysis for the energy-momentum-squared gravity, let's write the field equation (\ref{eq:14}) considering the energy-momentum tensor for the perfect fluid given in Eq. (\ref{eq:26}) and the expression for $\Theta_{\mu\nu}$ given in Eq. ({\ref{eq:28}}). For simplicity, it is chosen $f(R, T^2)=R+\lambda T^2$, where $\lambda$ is an integer. Then the Ricci tensor assumes the perfect fluid form, i.e.,
\bea
R_{\mu\nu}=ag_{\mu\nu}+b u_\mu u_\nu
\eea
with 
\bea
a&=& \frac{R}{2}+\frac{1}{2}\lambda T^2+\Lambda-\kappa p,\\
b&=&\kappa(\rho+p)+ f_{T ^{2}}(\rho^2-p^2).
\eea
This reinforces the idea that any term beyond the Ricci curvature scalar $R$ in the gravitational action can be modeled as a perfect fluid. This result does not change the discussions developed in previous sections about G\"{o}del-type universes in their behavior in $f(R, T^2)$ gravity.

\section{Conclusion}

The energy-momentum-squared gravity has been considered. This is a class of theories that generalize GR by adding higher order terms of the form $T_{\mu\nu}T^{\mu\nu}$ to the matter Lagrangian. In this context, the G\"{o}del and G\"{o}del-type solutions have been used to discuss the issue of causality and its violation. For a fluid perfect as the content of matter, it has been shown that the G\"{o}del metric is the solution of $f(R, T^2)$ gravity for specific values of the energy density $\rho$ and the cosmological constant $\Lambda$. This result leads to a violation of causality. At an appropriate limit, the GR results are recovered. For more details about this violation, the G\"{o}del-type is introduced. For simplicity, a local frame is used. In this context, different contents of matter are chosen. Using a perfect fluid as the matter content, the solution of the field equations shows that the violation of causality arises naturally. Here a finite critical radius, which explicitly depends on the function $f(R, T^2)$ and of matter content,  is calculated. The next investigation changes the matter content, now a perfect fluid plus a scalar field is considered. Our results lead to an infinite critical radius, therefore violation of causality is avoided. A similar result is obtained when only one scalar field is taken as matter content. Therefore, the energy-momentum-squared gravity allows G\"{o}del-types solutions, and as a consequence, both causal and non-causal regions emerge for different matter contents. Furthermore, the study developed in this paper is a generalization of the results obtained for $f(R)$ and $f(R, T)$ theories.  In addition, it is important to note that when the matter term is switched off, our results recover the results obtained by $f(R)$ gravity \cite{rebouccas2009godel}. In this case, the definition of causal and non-causal regions (which are given by the critical radius) continues to depend on the matter content that is introduced through the energy-momentum tensor.

\section*{Acknowledgments}

This work by A. F. S. is partially supported by National Council for Scientific and Technological Develo\-pment - CNPq project No. 313400/2020-2. A. J. C. C. thanks CAPES for financial support.


\global\long\def\link#1#2{\href{http://eudml.org/#1}{#2}}
 \global\long\def\doi#1#2{\href{http://dx.doi.org/#1}{#2}}
 \global\long\def\arXiv#1#2{\href{http://arxiv.org/abs/#1}{arXiv:#1 [#2]}}
 \global\long\def\arXivOld#1{\href{http://arxiv.org/abs/#1}{arXiv:#1}}



\begin{thebibliography}{99}

\bibitem{1916A} A. Einstein, ``Die Grundlage der allgemeinen Relativit\"{a}tstheorie'',
\doi{10.1002/andp.19163540702} {Annalen der Physik {354}, 769 (1916).}

\bibitem{einstein1915field} A. Einstein, ``The Field Equations of Gravitation''
 Sitzungsber. Preuss. Akad. Wiss. Berlin (Math.Phys.) {\bf 1915}, 844 (1915).

\bibitem{review1} C. M. Will, ``The Confrontation between General Relativity and Experiment'',
\doi{10.12942/lrr-2014-4} {Living Rev. Relativ. {\bf 17}, 4 (2014). }

\bibitem{review2} N. V. Krishnendu and F. Ohme, ``Testing General Relativity with Gravitational Waves: An Overview'', 
\doi{10.3390/universe7120497} {Universe {\bf 7} ,497 (2021).} 

\bibitem{Riess_1998} A. G. Riess, et al., ``Observational Evidence from Supernovae for an Accelerating Universe and a Cosmological Constant'',
\doi{10.1086/300499} {Astron. J. {\bf 116}, 1009 (1998).}

\bibitem{Perlmutter_1999} S. Perlmutter, et al.,``Observational Evidence from Supernovae for an Accelerating Universe and a Cosmological Constant'',
\doi{10.1086/307221} {Astron. J. {\bf 517}, 565 (1999).}

\bibitem{Riess_2004} A. G. Riess, et al., ``Type Ia Supernova Discoveries at z>1 From the Hubble Space Telescope: Evidence for Past Deceleration and Constraints on Dark Energy Evolution'',
\doi{10.1086/383612} {	Astrophys. J. {\bf 607}, 665 (2004).}

\bibitem{Carroll_2001} S. M. Carroll, ``The Cosmological Constant'',
\doi{10.12942/lrr-2001-1} {Living Rev. Relativ. {\bf 4}, 1 (2001).}

\bibitem{quinte} J. Martin, ``Quintessence:  a mini-review'',
\doi{10.1142/S0217732308027631} {Mod. Phys. Lett. A {\bf 23}, 1252 (2008).}

\bibitem{Capozziello_2006} S. Capozziello and S. Nojiri and S.D. Odintsov, ``Unified phantom cosmology: Inflation, dark energy and dark matter under the same standard'',
\doi{10.1016/j.physletb.2005.11.012} {Phys. Lett. B {\bf 632}, 597 (2006).}

\bibitem{taquions} T. Padmanabhan and T. Roy Choudhury, ``Can the clustered dark matter and the smooth dark energy arise from the same scalar field?'',
\doi{10.1103/PhysRevD.66.081301} {Phys. Rev. D {\bf 66}, 081301 (2002). }

\bibitem{gasChaplygin} M. C. Bento, O. Bertolami, and A. A. Sen, ``Generalized Chaplygin gas, accelerated expansion, and dark-energy-matter unification'',
\doi{10.1103/PhysRevD.66.043507} {Phys. Rev. D {\bf 66}, 043507 (2002).}

\bibitem{f(R)} T. P. Sotiriou and V. Faraoni, ``$f(R)$ theories of gravity'',
\doi{10.1103/RevModPhys.82.451} {Rev. Mod. Phys. {\bf 82}, 451 (2010).}

\bibitem{Nojiri} S. Nojiri, S.D. Odintsov, ``Unified cosmic history in modified gravity: from F(R) theory to Lorentz non-invariant models'',
\doi{10.1016/j.physrep.2011.04.001 } {Phys. Rep. {\bf 505}, 59 (2011).}


\bibitem{Carroll} S.M. Carroll, V. Duvvuri, M. Trodden, M.S. Turner,  ``Is Cosmic Speed-Up Due to New Gravitational Physics?''
\doi{10.1103/PhysRevD.70.043528} { Phys. Rev. D {\bf 70}, 043528 (2004).} 

\bibitem{Harko} T. Harko, F.S.N. Lobo, S. Nojiri and S.D. Odintsov, ``$f(R,T)$ gravity'',
\doi{10.1103/PhysRevD.84.024020} {Phys. Rev. D {\bf 84}, 024020 (2011).}

\bibitem{Clifton} T. Clifton, P. G. Ferreira, A. Padilla, C. Skordis, ``Modified gravity and cosmology'',
\doi{10.1016/j.physrep.2012.01.001} {Phys. Rep. {\bf 513}, 1 (2012).}

\bibitem{Shan} S. Shankaranarayanan, Joseph P Johnson, ``Modified theories of Gravity: Why, How and What?'',
\doi{10.1007/s10714-022-02927-2} {Gen. Relativ. Gravit. {\bf 54}, 44 (2022).}

\bibitem{Kat} N. Katirci and M. Kavuk, ``$f(R, T_{\mu\nu}T^{\mu\nu})$ gravity and Cardassian-like expansion as one
of its consequences'''
\doi{10.1140/epjp/i2014-14163-6} {Eur. Phys. J. Plus {\bf 129}, 163, (2014).}


\bibitem{Roshan} M. Roshan and F. Shojai, ``Energy-momentum squared gravity'',
\doi{10.1103/PhysRevD.94.044002} {Phys. Rev. D {\bf 94}, 044002 (2016).}

\bibitem{Barrow} C.V. R. Board and J. D. Barrow, ``Cosmological Models in Energy-Momentum-Squared Gravity'',
\doi{10.1103/PhysRevD.96.123517} {Phys. Rev. D {\bf 96}, 123517 (2017).}

\bibitem{Cosmic} \"{O}. Akarsu, N. Katirci and S. Kumar, ``Cosmic acceleration in a dust only universe via energy-momentum powered gravity'', 
\doi{10.1103/PhysRevD.97.024011} {Phys. Rev. D {\bf 97}, 024011 (2018).}


\bibitem{Uzun} \"{O}. Akarsu, J.D. Barrow, C.V.R. Board, N.M. Uzun and J.A. Vazquez, ``Screening $\Lambda$ in a new modified gravity model'', 
\doi{10.1140/epjc/s10052-019-7333-z} {Eur. Phys. J. C {\bf 79}, 846 (2019).}

\bibitem{Eksi} \"{O}. Akarsu, J.D. Barrow, S. Ckintoglu, K.Y. Eksi and N. Katirci, ``Constraint on energy-momentum squared gravity from neutron stars and its cosmological implications'', 
\doi{10.1103/PhysRevD.97.124017} {Phys. Rev. D {\bf 97}, 124017 (2018).} 

\bibitem{Nari} N. Nari and M. Roshan, ``Compact stars in energy-momentum squared'',
\doi{10.1103/PhysRevD.98.024031} {Phys. Rev. D {\bf 98}, 024031 (2018).}

\bibitem{Marciu} S. Bahamonde, M. Marciu and P. Rudra, ``Dynamical system analysis of generalized energy-momentum squared gravity'', 
\doi{10.1103/PhysRevD.100.083511} {Phys. Rev. D {\bf 100}, 083511 (2019).}

\bibitem{Barbar} A.H. Barbar, A.M. Awad and M.T. AlFiky, ``Viability of bouncing cosmology in energy-momentum squared gravity'', 
\doi{10.1103/PhysRevD.101.044058} {Phys. Rev. D {\bf 101}, 044058 (2020).}

\bibitem{Palatini} E. Nazari, F. Sarvi and M. Roshan, ``Generalized energy-momentum squared gravity in the Palatini formalism'', 
\doi{10.1103/PhysRevD.102.064016} {Phys. Rev. D {\bf 102}, 064016 (2020).} 

\bibitem{Modes} C.Y. Chen, M. Bouhmadi-Lopez and P. Chen, ``Lessons from black hole quasinormal modes in modified gravity'', 
\doi{10.1140/epjp/s13360-021-01227-z} {Eur. Phys. J. Plus {\bf 136}, 253 (2021).}

\bibitem{scale} O. Akarsu and N. M. Uzun, ``Cosmological models in scale-independent energy-momentum squared gravity'',
\doi{10.1016/j.dark.2023.101194} {Phys. Dark Univ. {\bf 40}, 101194 (2023).}

\bibitem{K.Godel} K.G\"{o}del, ``An Example of a New Type of Cosmological Solutions of Einstein's Field Equations of Gravitation',' 
\doi{10.1103/RevModPhys.21.447} {Rev. Mod. Phys. {\bf 21}, 447 (1949).}

\bibitem{Stockum} W. J. van Stockum,  ``The Precession of the Inertial Frame of a Rotating Body'',
\doi{} {Proceedings of the Royal Irish Academy. Section A: Mathematical and Physical Sciences {\bf  44}, 109 (1937)}

\bibitem{Kerr} R. P. Kerr, Phys. ``Gravitational Field of a Spinning Mass as an Example of Algebraically Special Metrics'',
\doi{10.1103/PhysRevLett.11.237} {Rev. Lett. {\bf 11}, 237 (1963).}

\bibitem{string} J. R. Gott, ``Closed timelike curves produced by pairs of moving cosmic strings: Exact solutions'',
\doi{10.1103/PhysRevLett.66.1126} {Phys. Rev. Lett. {\bf 66}, 1126 (1991).}

\bibitem{Godel-type.Reboucas} M. J. Rebou\c{c}as and J. Tiomno, ``Homogeneity of Riemannian space-times of G\"{o}del type','
\doi{10.1103/PhysRevD.28.1251} {Phys. Rev. D {\bf 28}, 1251 (1983).}

\bibitem{rebouccas2009godel} M. J. Rebou\c{c}as and J. Santos, ``G\"{o}del-type universes in $f({R})$ gravity'',
\doi{10.1103/PhysRevD.80.063009} {Phys. Rev. D {\bf 80}, 063009 (2009).}

\bibitem{Capo1} S. Capozziello, C. A. Mantica, and L. G. Molinari, ``Geometric perfect fluids from Extended Gravity'',
\doi{10.1209/0295-5075/ac525d} {EPL {\bf 137}, 19001 (2022).}

\bibitem{Capo2} S. Capozziello, C. A. Mantica, and L. G. Molinari, ``Cosmological perfect-fluids in $f(R)$ gravity'',
\doi{10.1142/S0219887819500087} {Int. J. Geom. Meth. Mod. Phys. {\bf 16}, 1950008 (2019).}

\bibitem{Capo3} S. Capozziello, C. A. Mantica, and L. G. Molinari, ``Cosmological perfect fluids in higher-order gravity'',
\doi{10.1007/s10714-020-02690-2} {Gen. Rel. Grav. {\bf 52}, 36 (2020).}

\bibitem{Sid} M. D. Siddiqi, S. K. Chaubey and M. N. I. Khan, ``$f(R, T)$-Gravity Model with Perfect Fluid Admitting Einstein Solitons'',
\doi{10.3390/math10010082} {Mathematics {\bf 10}, 82 (2022).}





\end{thebibliography}
\end{document}